\documentclass[12pt]{iopart}
 \usepackage{graphicx}

\begin{document}

\title{Enhanced Geometry Fluctuations in Minkowski and Black Hole Spacetimes}

\author{R T Thompson and L H Ford}

\address{Institute of Cosmology \\ 
Department of Physics and Astronomy \\
Tufts University, Medford, MA 02155}

\begin{abstract}
We will discuss selected physical effects of spacetime geometry
fluctuations, especially the operational signatures of geometry
fluctuations and their effects on black hole horizons. The 
 operational signatures which we discuss involve the effects of the
fluctuations on images, and include luminosity variations, spectral
line broadening and angular blurring. Our main interest will be in
black hole horizon fluctuations, especially horizon fluctuations
which have been enhanced above the vacuum level by gravitons or
matter in squeezed states. We investigate whether these fluctuations
can alter the thermal character of a black hole. We find that this
thermal character is remarkably robust, and that Hawking's original
derivation using transplanckian modes does not seem to be sensitive
 even to enhanced horizon fluctuations. 
\end{abstract}

\pacs{04.70.Dy,04.60.Bc,04.62.+v}

\maketitle

\section{Introduction}
\label{sec:intro}

In this paper, we will discuss selected aspects of the effects of
quantum spacetime geometry fluctuations, using the Riemann tensor 
correlation function as our basic tool. It is first useful to make 
a distinction between active fluctuations, which arise from the 
dynamical degrees of freedom of gravity itself, and passive
fluctuations, which are driven by quantum stress tensor fluctuations
of matter fields. In general, both types of fluctuations are
present. We will be concerned with the case where the fluctuations
around a classical background spacetime are small, so we can
consider each type separately and add their effects.  Of course, a
full treatment of active fluctuations would require a more complete
theory of quantum gravity than currently exists, but we restrict our
attention to quantized linear perturbations of the background
spacetime.

Various aspects of spacetime geometry fluctuations have been discussed
by several authors in recent years, for example
~\cite{F82,Kuo,CH95,PH97,CCV,HS98,MV99,WF99,NvD00,AC00,NCvD03,AC04}. 
In particular, stochastic gravity~\cite{Stochastic} has been studied 
as a systematic approach to go beyond the semiclassical theory.
In the present paper, we will concentrate on two issues. The first
is some phenomenological effects of geometry fluctuations in 
a nearly flat
background, which will be discussed in Section~\ref{sec:RTF}.
 Here we will be summarising work that was previously
published~\cite{BF,TF06}. The second issue will be the possibility of
enhancing horizon fluctuations by use of squeezed states, discussed in
Section~\ref{sec:enhanced}. Here we will be summarising work that
will developed in  more detail in a forthcoming paper~\cite{TF08}.

\section{Phenomenology of Riemann Tensor Fluctuations}
\label{sec:RTF}

In this section, we will briefly review some of the operational
signatures of spacetime geometry fluctuations, and describe how
these may be given a geometrical description in terms of the Riemann
tensor correlation function. Here we direct our attention to the case
of a nearly flat background, but the same techniques can be used
in more general spacetimes.

\subsection{Luminosity Fluctuations}

The image of a source viewed through a fluctuating medium can undergo
variations in apparent brightness. This effect is known to astronomers
as scintillation, and to the general public as ``twinkling''. The most 
familiar example arises when stars are viewed through the earth's
atmosphere, which is undergoing density fluctuations. In the case of 
fluctuations of the spacetime geometry, this effect may be studied
using the Raychaudhuri equation as a Langevin
equation~\cite{BF,Moffat}. 
One calculates
fluctuations of the expansion $\theta$. In the case that the shear,
vorticity, and squared expansion can be neglected, this equation
becomes
\begin{equation}
\frac{d \theta}{d \lambda}  = - R_{\mu\nu} k^\mu k^\nu \, ,
\end{equation}
where  $\lambda$ is an affine parameter, $k^\mu$ is the tangent to
the geodesic congruence, and   $ R_{\mu\nu}$ is the Ricci tensor.
If the expansion vanishes at $\lambda_0$, then the variance of the
expansion becomes
\begin{eqnarray}
\langle \theta^2 \rangle- \langle \theta \rangle^2 &=&
\langle (\Delta \theta)^2 \rangle=
\int_0^{\lambda_0} d\lambda \int_0^{\lambda_0} d\lambda' \nonumber \\
&\times& C_{\mu \nu \alpha \beta}(\lambda,\lambda')\, 
k^\mu (\lambda) k^\nu (\lambda)
\, k^\alpha(\lambda') k^\beta(\lambda') \,, \label{eq:var}
\end{eqnarray}
where 
\begin{equation}
C_{\mu \nu \alpha \beta}(x,x') = 
\langle R_{\mu \nu }(x) R_{\alpha \beta}(x') \rangle
- \langle R_{\mu \nu }(x) \rangle \langle R_{\alpha \beta}(x') \rangle 
\end{equation}
is the Ricci tensor correlation function. It is obtained by
appropriate contractions of the Riemann tensor correlation function and
is determined by the stress tensor correlation function. Thus we see
that luminosity fluctuations, in lowest order, are signatures of
passive fluctuations. The results of several explicit models are given 
in~\cite{BF}.

\subsection{Line Broadening and Angular Blurring}

\begin{figure}
  \centering
  \scalebox{.5}{\includegraphics{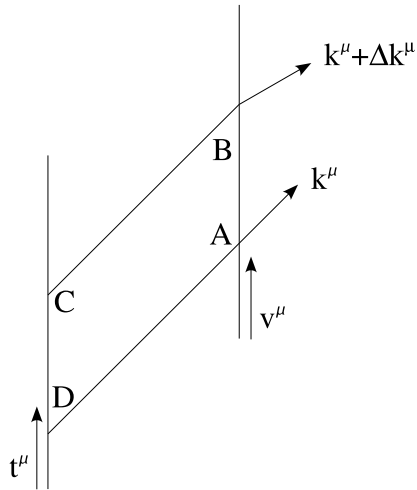}}\\
  \caption{A source moves along a worldline with tangent $t^{\mu}$
  while a detector a proper distance $s$ away moves along a worldline
  with tangent $v^{\mu}$. The source emits a ray  which has
  tangent $k^{\mu}(\lambda=0)$ at point D and tangent $k^{\mu}(\lambda_0)$ 
  at A.
  Parallel propagation of $k^{\mu}$
  around ABCD results in a slightly rotated vector $k^{\mu} + \Delta k^{\mu}$.
  The closed path ABCD encloses the spacetime region of interest.}
  \label{fig:spacetime}
\end{figure}

Two other effects which can arise when a source is viewed through
a fluctuating spacetime geometry are broadening of spectral lines and 
blurring of images. Both of these effects may be given a unified
geometric treatment, which we summarise here. Consider a source which is
sending successive pulses to a detector as illustrated in
Figure~\ref{fig:spacetime}.
The effect of the spacetime geometry is to cause  a shift $\Delta k^{\mu}$
between successive pulses. This shift can be expressed as an integral
of the Riemann tensor over the region illustrated:
\begin{equation}
 \Delta k^{\mu} = 
 - \int_{\tau_1}^{\tau_2} d\tau \int_0^{\lambda_0} d\lambda \,
 R^{\mu}_{\phantom{\mu}\alpha\nu\beta}k^{\alpha}t^{\nu}k^{\beta}\,.
\end{equation}
A shift in the time component is observed as a frequency shift,
whereas as shift in a spatial component produces a shift in position
of the image. Fluctuations of the geometry lead to variations in each
of these quantities. Denote the fractional line broadening by 
$\Delta \xi =\Delta \omega/\omega$. Its variance can be expressed
as
\begin{equation}
  \delta\xi^2 = \langle(\Delta\xi)^2\rangle -
  \langle\Delta\xi\rangle^2 =
  \int da \int da' \,
  C_{\alpha\beta\mu\nu\,\gamma\delta\rho\sigma}(x,x')
  t^{\alpha}k^{\beta}t^{\mu}k^{\nu}t^{\gamma}k^{\delta}t^{\rho}k^{\sigma}.
                    \label{eq:delta_xi}
\end{equation}
Here 
\begin{equation}\label{CorFunction}
 C_{\alpha\beta\mu\nu\,\gamma\delta\rho\sigma}(x,x') = \langle
 R_{\alpha\beta\mu\nu}(x)
 R_{\gamma\delta\rho\sigma}(x')\rangle -
 \langle R_{\alpha\beta\mu\nu}(x)\rangle
 \langle R_{\gamma\delta\rho\sigma}(x')\rangle
\end{equation}
is the Riemann tensor correlation function. Similarly the fluctuations
of an image's angular position in a direction defined by the spacelike
vector $s^\mu$ are given by.
 \begin{eqnarray}
 \delta\Theta^2 &=& \langle(\Delta\Theta)^2\rangle -
 \langle\Delta\Theta\rangle^2 = \int da\int da' \nonumber \\
&\times&  C_{\alpha\beta\mu\nu \, \gamma\delta\rho\sigma}(x,x')
 s^{\alpha}k^{\beta}t^{\mu}k^{\nu}s^{\gamma}k^{\delta}t^{\rho}k^{\sigma}.
                                    \label{eq:delta_theta}
\end{eqnarray}
Several explicit examples of these effects, including enhanced
fluctuations due to gravitons in squeezed states,
 are discussed in~\cite{TF06} .

\section{Black Hole Evaporation}

\begin{figure}
  \centering
  \scalebox{.3}{\includegraphics{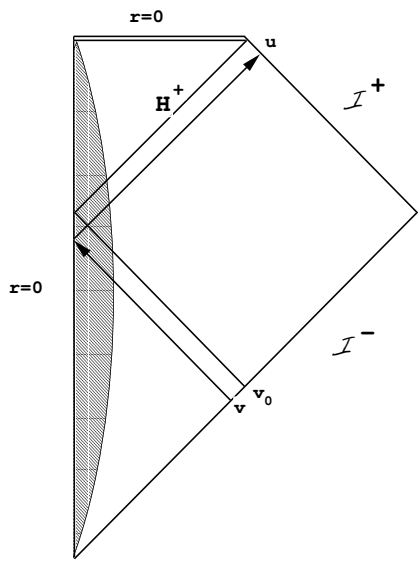}}\\
  \caption{The spacetime of a black hole formed by gravitational
    collapse is illustrated. The shaded region is the interior of the
collapsing body, the $r=0$ line on the left is the worldline of the center 
of this body, the $r=0$ line at the top of the diagram is the curvature
singularity, and $H^{+}$ is the future event horizon. An ingoing light
ray with $v < v_0$ from ${\cal I^{-}}$ passes through the body and escapes
to ${\cal I^{+}}$ as a $u = constant$ light ray. Ingoing rays with $v > v_0$
do not escape and eventually reach the singularity.}
  \label{fig:collapse}
\end{figure}
In this section, we will briefly review Hawking's
derivation~\cite{Hawking} of black hole
evaporation. The basic idea is to consider the spacetime of a black hole
formed by gravitational collapse, as illustrated in 
Figure~\ref{fig:collapse}. A quantum field
propagating in this spacetime is assumed to be in the in-vacuum state, that
is, containing no particles before the collapse.        In the case of a
massless field, a purely positive frequency mode proportional to
${\rm e}^{-i\omega v}$ leaves $\cal{I}^{-}$, propagates through the
collapsing body, and reaches $\cal{I}^{+}$ after undergoing a large redshift
in the region outside of the collapsing matter. At $\cal{I}^{+}$,
the mode is now a mixture of positive and negative frequency parts,
signalling quantum particle creation. Of special interest are the modes
which leave $\cal{I}^{-}$ just before the formation of the horizon, which is
the $v = v_0$ ray. These modes give the dominant contribution to the
outgoing flux at times long after the black hole has formed. After passing
through the collapsing body, they are $u$= constant rays, where
\begin{equation}
u = -4M\, \ln\biggl({{v_0 -v}\over C}\biggr)\, , \label{eq:uofv}
\end{equation}
where $M$ is the black hole's mass, and $C$ is a constant.
The logarithmic dependence leads to a Planckian spectrum of created
particles. It also leads to what some authors call the ``transplanckian
problem''. We prefer to refer it as the ``transplanckian issue'', as in our
view there may not be any problem. In any case, this is the enormous
frequency which the dominant modes must have when they leave $\cal{I}^{-}$.
The typical frequency of the radiated particles reaching $\cal{I}^{+}$
midway through the evaporation process is of order $1/M$, but the typical
frequency of these modes at $\cal{I}^{-}$ is of order
\begin{equation}
\omega \approx M^{-1} {\rm e}^{(M/M_{Pl})^2}\, ,
\end{equation}
 where $M_{Pl}$ is the Planck mass.
Another way to state this is to note that the characteristic value of $u$
for these modes is of order
\begin{equation}
u_c \approx M \biggl(\frac{M}{m_p}\biggr)^2 \,.
\end{equation}
A geodesic observer who falls from rest at large distance from the black
hole will pass from $u=u_c$ to the horizon at $u = \infty$ in a proper time
of
\begin{equation}
\delta \tau \approx M\, e^{-{u_c}/{4M}} \approx
                    M\, e^{-{M^2}/{m_p^2}}\,.   \label{eq:del-tau}              
\end{equation}
which is far smaller than the Planck time. In this sense, the outgoing modes
are much less than a Planck length outside the horizon.

Several authors, especially Unruh~\cite{Unruh1,Unruh2} and
Jacobson~\cite{Jacobson1,Jacobson2} have noted that it is
possible to avoid the use of transplanckian modes provided that the
dispersion relation for the quantum field is suitably modified. This can
lead to the phenomenon of ``mode regeneration'', whereby the modes which
become occupied by the thermal radiation do not arise by red-shifting of
transplanckian modes, but rather are generated by nonlinear effects just
before they are due to be occupied. This possibility has the benefit of
avoiding transplankian frequencies, but the drawback that it requires
violation of local Lorentz invariance, and hence would depend upon as yet
undiscovered physics.

For the case of a bosonic field, spontaneous emission implies the
possibility of stimulated emission. However, in the picture described by the
original Hawking derivation, stimulated emission into modes with wavelengths
of order $1/M$ require one to start with an initial state
which is populated by particles with transplanckian frequencies. The
possibility of stimulated emission in the case of the Hawking effect was
first discussed by Wald~\cite{Wald76}.

\section{Horizon Fluctuations}

We now turn to the central issue of this paper, that of black hole horizon
fluctuations. In classical general relativity, the event horizon is the
history of the light ray which is marginally trapped by the black hole.
It gives a sharp boundary between events which are visible to the outside,
and those which are not. However, in quantum physics, one does
not expect such sharp boundaries, but rather a smeared horizon due to
quantum fluctuation effects. Several authors have discussed this
possibility from different viewpoints
~\cite{Bekenstein,Sorkin1,Sorkin2,Casher,BFP00,FS97,Marolf,HR07}. One
approach 
is to calculate
fluctuations in the horizon area~\cite{Bekenstein,Sorkin1,Sorkin2}. 
Another~\cite{FS97} is to study the variation in
the times at which rays just outside the classical horizon arrive at
a distant observer. It is the latter approach which will adopted in
the present paper. 

A crucial question is whether quantum horizon fluctuations will
significantly alter the Hawking derivation outlined above. Given that
the outgoing modes which eventually become populated with the thermal
radiation spend part of their history less than a Planck length
outside the mean horizon, it would seem at first that horizon
fluctuations could drastically change the result. Fluctuations on the
Planck scale would seem to cause a significant fraction of the
outgoing modes either to fall into the black hole and never reach
$\cal{I}^{+}$, or else to be ejected prematurely. In either case, the
thermal character of the Hawking radiation would seem to be greatly
changed. This would upset the elegant connection between black hole 
physics, thermodynamics, and quantum theory.  In fact, the situation
is not so dire as it first seems. Ford and Svaiter~\cite{FS97} gave an estimate
of the effects of the active fluctuations when the graviton field is in 
its vacuum state. They found that the effects are quite small for
black holes whose mass is above the Planck scale. Specifically, they
calculated  variations in the time $\delta \tau$ given in
(\ref{eq:del-tau}) due to the quantum metric fluctuations and found
that these fractional variations are small so long as $M \gg m_{Pl}$.

\section{Enhanced Horizon Fluctuations}
\label{sec:enhanced}

Squeezed states are capable of enhancing 
quantum fluctuations beyond the vacuum level. This raises the
possibility of increasing the horizon fluctuations of a black hole
by sending in quantum fields in a squeezed vacuum state. In the
remainder of this paper, we will discuss three models for the source
of the fluctuations. However, first we outline a geometric approach
to study the effects of geometry fluctuations on the horizon. 
The basic idea is to study the geodesic deviation of the outgoing null
geodesics just outside the mean horizon. 
\begin{figure}
 \centering
 \scalebox{.5}{\includegraphics{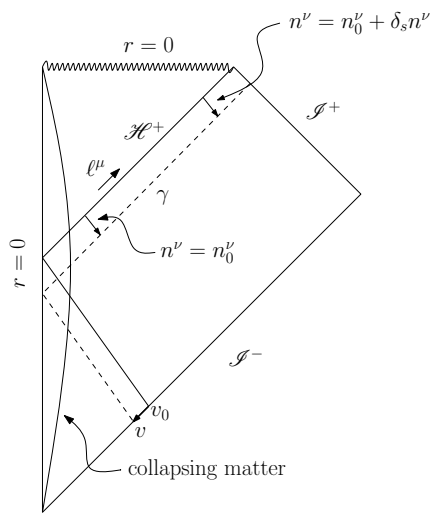}}\\
 \caption{Schwarzschild black hole space-time formed via gravitational 
collapse.  The separation vector characterizing geodesic deviation of
the 
horizon, $\cal{H}^+$, and a nearby outgoing null geodesic,
$\gamma$, 
is initially $n^{\nu}=n^{\nu}_0$ but then evolves as 
$n^{\nu}=n^{\nu}_0 + \delta_s n^{\mu}$.  The vector $\ell^{\mu}$ is 
tangent to the horizon.}
 \label{Fig:BlackHole2}
\end{figure}
This is illustrated in Figure~\ref{Fig:BlackHole2}. Here $n^\mu$ is
the separation vector which describes the peeling of the outgoing
geodesics from the horizon (which is hidden in a conformal spacetime
diagram). This vector satisfies the geodesic deviation equation,
\begin{equation}
 \frac{D^2n^{\alpha}}{dV^2} = 
R^{\alpha}_{\phantom{\alpha}\beta\mu\nu}\ell^{\beta}\ell^{\mu}
n^{\nu}\,,           \label{eq:geod}
\end{equation}
where $\ell^{\mu}$ is the tangent to the horizon.

We first consider the solution of this equation in the classical
Schwarzschild spacetime. This is conveniently done in null Kruskal
coordinates, in which the Schwarzschild metric is
\begin{equation}
 ds^2 = \frac{-32M^3}{r} e^{-r/2M} dU dV + r^2 d\Omega^2 \,.
\end{equation}
The advantage of these coordinates is that $V$, which is constant on
ingoing null rays, is an affine parameter for the horizon. Let $V=V_0$
be an initial point at which $n^\mu = n_0 \,\delta^\mu_U$, where $n_0$
is a constant. Then the solution of (\ref{eq:geod}) to lowest
non-trivial order in $V-V_0$, when $r-2M \ll M$, is
\begin{equation} \label{Eq:BackgroundDeviation}
 n^{\alpha} =  n_0^{\alpha}+\delta_s n^{\alpha} =
  n_0\left(1,-e^{-1}(V-V_0)^2\right) \,.
\end{equation}
That is, the vector $n^\mu$ which initially has only a $U$-component,
develops a $V$-component with no change in the $U$-component to this
order. The vector is null at the starting point, but afterwards is 
spacelike with squared norm 
\begin{equation}
 \left(n^{\mu}n_{\mu}\right)^2 = 
4\left(g_{UV}\, n_0^2 e^{-1}\right)^2\left(V-V_0\right)^4 \,.
\end{equation}
The growth of $n^V$ describes the peeling of the outgoing geodesic
from the horizon. 

We now wish to consider the effects of quantum geometry fluctuations.
This may be done by treating (\ref{eq:geod}) as a Langevin equation
with a fluctuating part of the Riemann tensor, 
 $\delta R^{\alpha}_{\phantom{\alpha}\mu\beta\nu}$. The separation
vector now becomes a fluctuating vector written as
\begin{equation} \label{Eq:SeparationEvolution}
 \bar n^{\alpha} = n_0^{\alpha}+\delta_s n^{\alpha} + 
\delta_p \hat n^\alpha\,,
\end{equation}
where the contribution from the fluctuations, to lowest order in
$V-V_0$, is
\begin{equation}
\delta_p \hat n^\alpha =
\int_{V_0}^{V} dW\int_{V_0}^{W}dV\, 
\delta
R^{\alpha}_{\phantom{\alpha}\mu\beta\nu}\ell^{\mu}\ell^{\beta}n_0^{\nu}.
   \label{eq:del_p}
\end{equation}
Our measure of the effects of the quantum fluctuations will be the
variance of $(\bar n^{\mu}\bar n_{\mu})^2$. To lowest order in
$V-V_0$, this variance may be expressed as
\begin{eqnarray} \label{Eq:variance}
 \Delta(\bar n^{\mu}\bar n_{\mu})^2 &=& 
4\left[\langle \left(n^{\mu}_0(x)\delta_p \hat n_{\mu}(x)\right)
\left(n^{\nu}_0(x')\delta_p \hat n_{\nu}(x')\right)\rangle \right.
\nonumber  \\ 
 &-& \left.  \langle n^{\mu}_0(x)\delta_p \hat n_{\mu}(x)\rangle
\langle n^{\nu}_0(x')\delta_p \hat n_{\nu}(x')\rangle\right].
\end{eqnarray}
 Here we have used the facts that $\delta_s n_{\nu}(x)\propto
 (V-V_0)^2$, and that  each $\delta_p \hat n_{\nu}(x)\propto
 (V-V_0)^2$, as will be demonstrated in each of the explicit models
of fluctuations to be considered below. Note that (\ref{Eq:variance})
involves a double integral of the Riemann tensor correlation function,
as each factor of $\delta_p \hat n_{\nu}$ is an integral over the
fluctuating Riemann tensor, (\ref{eq:del_p}).

\subsection{Scalar Graviton Model} 
\label{sec:ScalarGraviton}

First we discuss a simplified model which captures the essential
physics of active fluctuations without the complications entailed in
treating tensor perturbations in a black hole spacetime. Here the
perturbed metric is taken to have the form
\begin{equation}
\bar g_{\mu\nu} = \left( 1+\Phi\right)g_{\mu\nu} \,,
\end{equation}
where $g_{\mu\nu}$ is the unperturbed Schwarzschild background  metric
and $\Phi$ is a free quantum scalar field, which we take to satisfy
the minimally coupled wave equation
\begin{equation}
\nabla^\mu \nabla_\mu\, \Phi =0
\end{equation}
in the Schwarzschild spacetime. This is sometimes called a dilaton
field, but for our purposes it models the effects of gravitons. 
We can expand the operator $\Phi$ in terms of a complete set of
positive norm wave packets solutions $\{\psi_{jn}\}$  as
\begin{equation}
 \Phi = \sum_{jn} (\psi_{jn} \hat{a}_{jn} + \psi_{jn}^*
 \hat{a}_{jn}^{\dagger}) \,,
                                   \label{eq:Phi}
\end{equation}
where the $ \hat{a}_{jn}$ are annihilation operators. 

Following
Hawking~\cite{Hawking}, we define the ingoing wave packets by
\begin{equation} \label{Eq:WavePacket}
 \psi_{j n} = \varepsilon_j^{-\frac{1}{2}} 
\int_{j\varepsilon_j}^{(j+1)\varepsilon_j} 
e^{-2\pi i n \omega/\varepsilon_j} \psi_{\omega \ell m} \, d\omega\,,
\end{equation}
where
\begin{equation}
 \psi_{\omega \ell m} = \frac{Y_{\ell m}(\theta,\varphi)}{r\sqrt{2\pi\omega}} 
F_{\omega}(r) e^{-i\omega v} \,. \label{eq:monochromatic}   
\end{equation}
Here $v = t +r^*$ is the usual advanced time coordinate. The function
$F(r)$ approaches unity for $r \gg M$ and approaches a constant of magnitude
less than unity, the transmission coefficient, near the horizon.  
The integer $j$ controls where in 
frequency space the wave packet is peaked, while $\varepsilon_j$
controls the width of the wave packet and has units of frequency.  
The integer $n$ describes which wave packet is under consideration.  
This construction allows for wave packets to be sent in at regular
intervals 
of ${2\pi}/{\varepsilon_j}$ with various frequencies.  
Thus $\psi_{jn}$ is the $n^{th}$ wave packet sent in with component 
frequencies ranging from $j\varepsilon_j$ to $(j+1)\varepsilon_j$.
A corresponding set of outgoing packets can be defined by substitution
of $u$ for $v$ in (\ref{eq:monochromatic}). 

We now wish to take the quantum state of the scalar field to be a
multi-mode squeezed vacuum state for some finite set of wave packet
modes, which are sent into the black hole well after the collapse.
Let the squeeze parameter of a given mode be
 $\xi_j = \rho_j\, {\rm e}^{i\delta_j}$, where $j$ labels the various
excited modes. We are interested in the effects of the excitation, so
we take the difference between the given state and the vacuum state,
and also assume that $\rho_j \gg 1$, corresponding to highly excited 
states. The function $F(r)$ becomes the transmission coefficient near
the horizon, where we need to evaluate it. This transmission coefficient
is of order unity for wavelengths shorter than the size of the black
hole, $\omega \geq (2M)^{-1}$, and is approximately $C (2i\omega
M)^{\ell+1}$ when $\omega \ll (2M)^{-1}$, where $|C|$ is a constant
of order unity.

In the case that  $\omega \approx (2M)^{-1}$, we find the fractional
fluctuations to be approximately
\begin{eqnarray}
 \frac{\Delta(\bar n^{\mu}\bar n_{\mu})^2}{(n^{\mu}n_{\mu})^2} &=& 
\frac{Y^2_{\ell 0}}{16}\, |C|^2 \, \sum_{j} 
\frac{e^{2\rho_j}\ell^2_P}{M\pi\varepsilon_j} \nonumber \\
&\times& \left[\int_{j\varepsilon_j}^{(j+1)\varepsilon_j} d\omega \, 
\sin\omega(v_0+\delta_j)+ 2\cos\omega(v_0+\delta_j)\right]^2\,,
\end{eqnarray}
where $\ell_P$ is the Planck length. 
The corresponding result for the low frequency limit, 
$\omega \ll(2M)^{-1}$, is
\begin{eqnarray}
 \frac{\Delta(\bar n^{\mu}\bar n_{\mu})^2}{(n^{\mu}n_{\mu})^2} &=& 
\frac{Y^2_{\ell 0}}{8}\sum_{j} \frac{e^{2\rho_j} 
\ell^2_P (2M)^{2\ell}}{\pi\varepsilon_j}   \nonumber \\
&\times&   \left[\int_{j\varepsilon_j}^{(j+1)\varepsilon_j} d\omega \, 
\omega^{\ell+1/2} \cos\omega(v_0+\delta_j)\right]^2.
\end{eqnarray}
These results may be further simplified to yield the estimates
\begin{equation}
 \frac{\Delta(\bar n^{\mu}\bar n_{\mu})^2}{(n^{\mu}n_{\mu})^2} \approx
 \sum_j \frac{e^{2\rho_j}\ell_P^2 \Delta\omega}{\pi M},
\qquad \omega \approx (2M)^{-1} \, ,  \label{eq:SG1}
\end{equation}
and 
\begin{equation}
 \frac{\Delta(\bar n^{\mu}\bar n_{\mu})^2}{(n^{\mu}n_{\mu})^2} \approx
 \sum_j \frac{1}{\pi}e^{2\rho_j}\ell_P^2(2M\omega)^{2\ell} \omega 
\Delta\omega, \qquad \omega \ll(2M)^{-1} \,.  \label{eq:SG2}
\end{equation}
Here $\Delta \omega$ is the bandwidth, a characteristic value of the 
$\varepsilon_j$, and we assume that $\Delta \omega \ll \omega$, where
$\omega$ is the typical peak frequency of the packets.

In principle, we can make $\rho_j$ and hence the fractional
fluctuations arbitrarily large. However, we have assumed that the
average background spacetime is close to Schwarzschild. This 
assumption will fail to be valid once the perturbation of the Riemann
tensor becomes of the same order as the background Riemann tensor.
This occurs when
\begin{equation}
\sum_j \frac{e^{2\rho_j}\ell_P^2 \vert F(\omega)\vert^2 
\Delta\omega}{M(M\omega)} \approx 1 \,. \label{eq:SGlimit} 
\end{equation}
If we combine this result with (\ref{eq:SG1}), we see that the
back reaction becomes large when the fractional fluctuations in 
$(\bar n^{\mu}\bar n_{\mu})^2$ are of order unity. This is expected,
as these fractional fluctuations are determined by the  fractional
change in curvature.

\subsection{Graviton Model} 
\label{sec:Graviton}

In this section, we briefly outline a more realistic model involving
gravitons propagating on the Schwazlrschild background. The gravitons
are quantized perturbations of the Schwarzschild geometry. Classical
black hole perturbations have been discussed by many authors,
beginning with Regge and Wheeler~\cite{RW}. We adopt the Regge-Wheeler gauge
explicit calculations, but as our key object is the gauge invariant
Riemann tensor correlation function, the results do not depend upon
this choice. As in the previous section, we consider graviton
wave packets in squeezed vacuum states which are sent into the black
hole after the collapse. The details of the calculation are
considerably more complicated than in the case of the scalar graviton
model, but the results~\cite{TF08} are essentially of the same
order. Specifically, both (\ref{eq:SG1}), (\ref{eq:SG1}), and 
(\ref{eq:SGlimit}) still apply to the graviton model. Again, one can
have the fractional fluctuations in 
$(\bar n^{\mu}\bar n_{\mu})^2$  of order unity.

\subsection{Passive Fluctuation Model} 
\label{sec:Passive}

In this model, the geometry fluctuations arise from the quantum stress
tensor fluctuations of a massless, minimally coupled scalar field. In
general, the stress tensor correlation function, and hence the    
Riemann tensor correlation function, are singular in the coincidence
limit even if the expectation value of the stress tensor has been 
renormalized. Specifically, one can decompose the stress tensor
correlation function into a sum of a fully normal-ordered part, a
cross term and a vacuum term. The latter two terms are singular 
in the coincidence limit, but can be defined as distributions by an
integration by parts procedure~\cite{WF01}, or by dimensional 
regularization~\cite{FW04}. For our purposes, it is sufficient to
include only the fully normal-ordered term, as this term gives the 
dominant contribution in the limit of highly excited states. In this
limit, the approximation of keeping only the fully normal-ordered term
is equivalent to that of including only those modes which are
excited.

Perturbation of the Riemann tensor comes entirely from Ricci tensor
parts in this model and is given by 
\begin{equation}
\delta R_{\alpha\beta\mu\nu} = 
8\pi\left[g_{\alpha[\mu}T_{\nu]\beta}-g_{\beta[\mu}T_{\nu]\alpha}- 
\frac{2}{3} g_{\alpha[\mu}g_{\nu]\beta}T\right] \,,
\end{equation}
where the scalar field stress tensor is 
\begin{equation}
 T_{\mu\nu}=\Phi_{;(\mu}\Phi_{;\nu)}-
\frac{1}{2}g_{\mu\nu}g^{\sigma\rho}\Phi_{;(\sigma}\Phi_{;\rho)}\,.
\end{equation}
The brackets denote anti-symmetrisation and the parentheses
symmetrisation. The scalar field operator $\Phi$ is given by
(\ref{eq:Phi}) and (\ref{Eq:WavePacket}).

The results of lengthy calculations for the fractional fluctuations
are
\begin{eqnarray}
 \frac{\Delta(\bar n^{\mu}\bar n_{\mu})^2}{(n^{\mu}n_{\mu})^2} &=& 
\left(\frac{16}{3}\right)^2 \, |C|^4 \, \sum_{j,k}
\frac{e^{2\rho_j} \ell_P^2 M}{\pi\varepsilon_j} \nonumber \\
&\times& \left[ \int_{j\varepsilon_j}^{(j+1)\varepsilon_j}d\omega\,
 \omega^{3/2}\cos\omega(v_0+\delta_j)\right]^2  \nonumber \\ 
&\times& \frac{e^{2\rho_k} \ell_P^2 M}{\pi\varepsilon_k} 
\left[ \int_{k\varepsilon_k}^{(k+1)\varepsilon_k}d\omega'\,
 (\omega')^{1/2}\cos\omega'(v_0+\delta_k)\right]^2 \,,
\end{eqnarray}
for the case $\omega \ll 1/M$, and
\begin{eqnarray}
 \frac{\Delta(\bar n^{\mu}\bar n_{\mu})^2}{(n^{\mu}n_{\mu})^2} &=& 
\left(\frac{4}{3}\right)^2 \sum_{j,k} 
\frac{e^{2\rho_j} \ell_P^2}{\pi \varepsilon_j M} 
\left[ \int_{j\varepsilon_j}^{(j+1)\varepsilon_j}
d\omega\,\sin\omega  v_0 \right]^2 \nonumber \\ 
&\times& \frac{e^{2\rho_k} \ell_P^2}{\pi \varepsilon_k M} 
\left[ \int_{k\varepsilon_k}^{(k+1)\varepsilon_k}d\omega'\,
  \sin\omega' v_0
\right]^2\,,
\end{eqnarray}
for the case $\omega \approx 1/M$. Note that in contrast to the scalar
graviton and graviton models, we now have a double sum over the
excited wave packet modes. This arises from the fact that the Riemann
tensor correlation function is quartic in the field operators.
We can estimate the fractional fluctuations for the case 
$\omega \ll1/M$ as
\begin{equation}
 \frac{\Delta(\bar n^{\mu}\bar n_{\mu})^2}{(n^{\mu}n_{\mu})^2} \approx
\left( \sum_j \frac{e^{2\rho_j}\ell_P^2 \Delta\omega}{\pi M} \right)^2\,.
\end{equation}

Comparison with (\ref{eq:SG1}), which hold for both the scalar graviton
and graviton models, reveals that the active fluctuations are
proportional to $\ell_P^2$, whereas the passive fluctuations are
proportional to  $\ell_P^4$. In this sense, active fluctuations tend
to dominate over passive ones. However, the passive fluctuations grow
more rapidly for large $\rho_j$. The point at which the mean
background spacetime ceases to be Schwarzschild is again when 
\begin{equation}
\frac{\Delta(\bar n^{\mu}\bar n_{\mu})^2}{(n^{\mu}n_{\mu})^2} \approx 1\,.
\end{equation}

\section{Summary and Discussion}
\label{sec:final}

We have discussed the problem of spacetime geometry fluctuations using
the Riemann tensor correlation function, first near flat spacetime,
and then in black hole spacetime. In the former case, we noted some of
the operational signatures of spacetime geometry fluctuation: 
luminosity fluctuations, line broadening, and angular
blurring. However, the main interest is in the problem of black hole
horizon fluctuations. 

The Hawking derivation of black hole radiance invokes transplanckian
modes, which must remain extremely close to the event horizon for
a very long time, as measured by an external observer. Here extremely
close means far less than one Planck length as measured by an
infalling observer. This suggests that quantum fluctuations of the
horizon might drastically alter black hole radiance. If this were the
case, then the connection between black hole physics and thermodynamics
might only be preserved by going to a ``mode-regeneration'' picture
based upon a non-linear, non-Lorentz invariant dispersion relation.
However, a previous study~\cite{FS97} 
of active fluctuations from gravitons in a
vacuum state indicated that the vacuum level horizon fluctuations will  
not upset the Hawking derivation for black holes more massive than the 
Planck mass.

In the present paper, we have summarised an investigation~\cite{TF08} 
which goes
further and considers both active and passive fluctuations from
squeezed states. This has the advantage that the level of the
fluctuations can be increased by increasing the squeeze parameter.
Indeed, we did find that one can the fractional fluctuation in the
geodesic deviation vector approach order unity. However, in all
of the models discussed in this paper, the contribution to the 
quantity ${\Delta(\bar n^{\mu}\bar n_{\mu})^2}/{(n^{\mu}n_{\mu})^2}$
come only from the $V$-component of $n^\mu$, not the $U$-component.
If we refer to Figure~\ref{Fig:BlackHole2} , 
we see that it would require $U$-component
fluctuations to cause serious problems with modes being prematurely
ejected or captured by the black hole. Fluctuations in the
$V$-component do influence when a given wave packet arrives at
$\cal{I}^+$. However, this is likely to be unobservable when the  
quantum state before collapse is the vacuum state. These fluctuations
could alter stimulated emission, and hence are in principle
observable, but would require a state containing particles with 
transplanckian energies if the stimulated emission is to be observed
well after the collapse.

In summary, although it is possible to enhance spacetime geometry
fluctuations by use of squeezed states, these enhanced fluctuations
do not seem to alter the thermal Hawking radiation. Thus we conclude
that Hawking's derivation of  black hole radiance is remarkably
robust, and that there may be no transplanckian ``problem''.

 \section*{Acknowledgment}
  This work was supported in part by the National
Science Foundation under Grant  PHY-0555754.

\end{document}